\documentclass[twocolumn]{aastex62}
\usepackage{graphics,epsf}
\usepackage{amsmath}                
\usepackage{amsfonts}               
\usepackage{amssymb}                
\usepackage{epsfig}                 
\usepackage{appendix}
\usepackage{graphicx}
\usepackage{float}
\usepackage{color}
\usepackage[para,online,flushleft]{threeparttable}

\newcommand{\kms}{{~\rm km\; s^{-1}}}

\newcommand{\cm}{{~\rm cm}}

\newcommand{\km}{{~\rm km}}
\newcommand{\s}{{~\rm s}}

\newcommand{\g}{{~\rm g}}

\newcommand{\K}{{~\rm K}}
\newcommand{\erg}{{~\rm erg}}
\newcommand{\yr}{{~\rm yr}}
\newcommand{\myr}{{~\rm Myr}}

\newcommand{\kpc}{{~\rm kpc}}

\begin{document}

\title{Kinematics of filaments in cooling flow clusters and heating by mixing}

\email{soker@physics.technion.ac.il}

\author{Shlomi Hillel}
\affiliation{Department of Physics, Technion – Israel Institute of Technology, Haifa 3200003, Israel}

\author{Noam Soker}
\affiliation{Department of Physics, Technion – Israel Institute of Technology, Haifa 3200003, Israel}
\affiliation{Guangdong Technion Israel Institute of Technology, Guangdong Province, Shantou 515069, China}

\begin{abstract}
We compare a recent study of the kinematics of optical filaments in three cooling flow clusters of galaxies with previous numerical simulations of jet-inflated hot bubbles, and conclude that the velocity structure functions of the filaments better fit direct excitation by the jets than by turbulent cascade from the largest turbulent eddies. 
The observed velocity structure functions of the optical filaments in the three clusters are steeper than that expected from a classical cascade in turbulent dissipation. Our three-dimensional (3D) hydrodynamical simulations show that as the jets inflate bubbles in the intracluster medium (ICM), they form vortexes in a large range of scales. These vortexes might drive the ICM turbulence with eddies of over more than an order of magnitude in size. A direct excitation of turbulence by the vortexes that the jets form and the slow turbulent dissipation imply that heating the ICM by mixing with hot bubbles is more efficient than heating by turbulent dissipation. 
\end{abstract}

  
\section{INTRODUCTION}
\label{sec:intro}

In the cooling flow process in clusters of galaxies, in galaxies, and during galaxy formation, the gas radiative cooling time is shorter than the age of the system. The thermal state of the intracluster medium (ICM; or the interstellar medium, ISM) is determined by radiative cooling and by heating, both operate together in a negative feedback cycle (for reviews, e.g., \citealt{Fabian2012, McNamaraNulsen2012, Soker2016, Werneretal2019}). In one direction of the feedback cycle the gas suffers radiative cooling and feeds the active galactic nucleus (AGN), while in the other direction jets that the central AGN launches heat the gas (e.g., \citealt{Farage2012, Gasparietal2013a, Pfrommer2013, Baraietal2016, Soker2016, Birzanetal2017, Iqbaletal2017, Wangetal2019}).
  
Many studies in recent years support the cold feedback mechanism \citep{PizzolatoSoker2005, Gasparietal2013b, Voitetal2015Natur} by which cold clumps that radiatively cool from the hot ICM (or from the hot ISM) feed the AGN (e.g., a small sample of papers from past 3 years, \citealt{Davidetal2017, Donahueetal2017, FujitaNagai2017, Gasparietal2017, Hoganetal2017, Prasadetal2017,  Babyketal2018, Gasparietal2018, Jietal2018, Prasadetal2018, Pulidoetal2018, Voit2018Turb, YangLetal2018, Choudhuryetal2019, Ianietal2019, Qiuetal2019, Roseetal2019, Russelletal2019, Sternetal2019, StorchiBergmann2019, Vantyghemetal2019, Voit2019}). 

On the other hand, there is an ongoing dispute on the main heating processes of the ICM. We list the several different heating processes as follows. (1) \textit{Cosmic rays} that are accelerated within the jet-inflated bubbles, stream into the ICM and heat it (e.g. \citealt{FujitaOhira2013, Pfrommer2013, Ehlertetal2018, Ruszkowskietal2018}). However, it seems that even in the case where the jet-inflated bubbles are filled with cosmic rays, mixing of the bubble content with the ICM (the heating by mixing process; see below) is more efficient than streaming of cosmic rays along magnetic field lines \citep{Soker2019CR}. (2) Excitation of \textit{sound waves} in the ICM (e.g., \citealt{Fabianetal2006, TangChurazov2018}). (3) Driving \textit{shocks} in to the ICM (e.g., \citealt{Randalletal2015, Guoetal2018}). (4) Powering \textit{turbulence} (e.g.,  \citealt{DeYoung2010, BanerjeeSharma2014, Gasparietal2014, Zhuravlevaetal2018}). (5) \textit{Uplifting gas} from inner regions (e.g., \citealt{GendronMarsolaisetal2017}). {{{{ (6) Generation of \textit{internal waves} in the ICM by buoyantly rising bubbles (e.g., \citealt{Zhangetal2018}). }}}}  (7) \textit{Heating by mixing} that operates through the many vortexes that the jets form as they interact with the ICM and inflate the bubbles. These vortexes mix the ICM with the energetic content of the bubbles (whether cosmic rays or thermal hot gas), and by that heat the ICM (e.g., \citealt{BruggenKaiser2002, Bruggenetal2009, GilkisSoker2012, HillelSoker2014, YangReynolds2016b}). The changing of the jets' axis direction with time allows efficient mixing of the entire inner ICM volume, (e.g., \citealt{Soker2018RNAAS, Cieloetal2018}).  

{{{{ Our past simulations show that it is possible that only part of the hot bubble gas mixes with the ICM and heats it. The other part of the hot bubble, mainly the inner part, continues to buoy out through the cluster and forms a rising bubble that is no longer powered by a jet. Such outer bubbles (outer X-ray cavities) are observed in a number of clusters  (e.g.,  \citealt{Fabianetal2011, Randalletal2015}).  \cite{GilkisSoker2012} conduct a hydrodynamical simulation of jet-inflated bubbles and find that mixing is the main heating mechanism. Although the jet (they simulate only one half of the space) was active for only the first 20~Myr of the simulation, a well-defined bubble is still rising through the ICM at $t=80~$Myr, when the bubble is at about 30~kpc from the center (their figure 5). In \cite{HillelSoker2017Gentle} we further analysed our simulations of heating by mixing \citep{HillelSoker2016}, and found that alongside with the heating by mixing the bubbles maintain their identity with unmixed hot gas, up to tens of kpc from the center (our grid was up to 50~kpc). We can see a hot low-density bubble at a distance of about 40~kpc from the center at $t=80~$Myr (see their figure 2; the jet was periodically active, therefore, other bubbles trail the first bubble). }}}}

There are several recent studies of the turbulent motion in the ICM, observationally (e.g., \citealt{Hitomi2016, Hitomi2018, Simionescuetal2019, Sandersetal2020}) and numerically (e.g., \citealt{MiniatiBeresnyak2015, Vazzaetal2017, Vazzaetal2018, Yangetal2019ApJ}). In a recent study \cite{Fujitaetal2020} suggest that the heating by mixing works, but that most mixing is by ICM turbulence that is formed by continuous accretion of gas onto the cluster. Our simulations (e.g., \citealt{GilkisSoker2012, HillelSoker2016}) show that jet-excited turbulence is sufficient to induce the required mixing.

In another recent paper \cite{Lietal2020} study the kinematics of optical filaments in the cooling flow clusters Perseus, Abell~2597 and Virgo, and find the motion of filaments to be turbulent (section \ref{sec:Turbulence}). They further conclude that their result is consistent with turbulence as an important heating mechanism, {{{{ supporting earlier claims from the results of Hitomi \citep{Hitomi2016}. }}}}
In this paper we present an opposite view. Despite that turbulence is present in the ICM (e.g., \citealt{Zhuravlevaetal2014, AndersonSunyaev2016, Arevalo2016, Hofmannetal2016, Zhuravlevaeta2019}), and might play a role in the evolution of condensations in the cold feedback mechanism (e.g., \citealt{Voit2018Turb}), some  studies find heating by turbulence to have limited efficiency (e.g., \citealt{Falcetaetal2010, Reynoldsetal2015, Hitomi2016, Bambicetal2018, MohapatraSharma2019, Valdarnini2019}). {{{{ However, many of these do not recover the directly-measured velocity with Hitomi, or have other  inconsistencies with observations. \cite{Reynoldsetal2015} simulate violent feedback rather than a gentle feedback as observations suggest \citep{Hoganetal2017},  \cite{Bambicetal2018} assume that turbulence is generated in the cluster center rather than by bubbles in a large volume of the core, and \cite{ MohapatraSharma2019} ignore the effects of stratification in the ICM. In that respect we note that we have shown that the simulations we performed in 2016 \citep{HillelSoker2016} and that we further analyse in this study, can account for the observations of Hitomi \citep{HillelSoker2017b}, and yield a gentle heating  \citep{HillelSoker2017Gentle}.  }}}}

In the present study we present our view (sections \ref{sec:Turbulence}, \ref{sec:flow}, and \ref{sec:VSF}) that the heating by mixing process better fits the new findings of \cite{Lietal2020}. We summarise in section \ref{sec:summary}.

\section{FILAMENT KINEMATICS}
\label{sec:Turbulence}

\cite{Lietal2020} analyse optical observations of filaments in three cooling flow clusters. They pair many different regions and record the velocity difference within each pair $\vert \delta v \vert$, and bin the different pairs by the distance $L$ between the two regions of each pair. They calculate the average absolute value of the velocity differences within each distance bin, $V_p (L)\equiv \langle \vert \delta v \vert \rangle$. The function $V_p(L)$ is the velocity structure function (VSF) of the optical filaments.

\cite{Lietal2020} conclude that on small scales $L<L_m$, where $L_m$ is the scale of the driving force, which they calculate from the typical size of the jet-inflated bubbles in each cluster, the velocity structure function is steeper than the classical Kolmogorov expectation. 
They infer that the turbulent driving scales of the three clusters are $L_m {\rm (Perseus)} \approx 10 \kpc$, $L_m {\rm (A2597)} \approx 4 \kpc$, and $L_m {\rm(Virgo)} \approx  1-2 \kpc$. 

From figure 2 of \cite{Lietal2020} we approximate the velocity structure function for $L< L_m$ by a power law, $V_p \propto L^k$. These approximate velocity structure functions for small scales in the three clusters are  
\begin{equation}
Vp {\rm (Perseus)} \propto L^{0.5}_p, \qquad 0.3  \kpc \la L \la 7 \kpc,  
\label{eq:Perseus}
\end{equation}
\begin{equation}
Vp {\rm (A2597)} \propto L^{0.8}_p, \qquad 0.3 \kpc \la L \la 4 \kpc,  
\label{eq:A2507}
\end{equation}
and
\begin{equation}
Vp {\rm (Virgo)} \propto L^{0.9}_p, \qquad 0.2 \kpc \la L \la 3 \kpc.  
\label{eq:Virgo}
\end{equation}

These functions teach us two important things. 
The first, as \cite{Lietal2020} notice, is that if there is no dissipation on all scales, these steeper-than-Kolmogorov velocity structure functions imply that the energy dissipation of the turbulence is much below the value that the large scale gives $Q_m \approx \rho V^3_p(L_m) /L_m$.  We note that $Q_m$ is already short of explaining heating in Perseus \citep{Hitomi2016}. While in the Kolmogorov velocity structure function the contribution of each scale is the same down to the dissipation length, for the three velocity structure functions above the contribution to heating, $Q(L)  \propto L^{3k-1}$, rapidly decreases for shorter scales since $3k-1 =0.5$, $1.4$, and $1.7$, for the three clusters, respectively.  

The second property that these velocity structure functions reveal is that the dissipation time is longer than the time between consecutive jet-launching episodes in these clusters. The dissipation time is few times the turnover time $t_L \simeq L/V_p(L)$. 
\cite{Lietal2020} notice this for the largest scale, and here we emphasise this also for the smaller scales.
For example, in Perseus they take $V_p(L_m) \simeq 140 \km \s^{-1}$, which gives $t_m(10 \kpc) \approx 70 \myr$ and a dissipation time of $t_{\rm diss}(10 \kpc) > 100 \myr$. The period of AGN activity in Perseus is {{{{ (\citealt{Lietal2020}; highly uncertain) }}}} $t_{\rm AGN} \approx 10 \myr \ll t_{\rm diss}$. Even for the smallest scale in Perseus the turnover time is longer than the jet activity cycle, 
$t_m (0.2 \kpc) \approx 13 \myr$, implying a dissipation time of $t_{\rm diss}(0.2 \kpc) >20 \myr$. 
{{{{ The inequality $t_{\rm AGN} \ll t_{\rm diss}$ implies that {{{{{ over a limited span of time that is not much longer than $t_{\rm diss}$, }}}}} the turbulence can transfer only a small fraction of the AGN power to heat the ICM. Since in many clusters the power of AGN activity is about equal or not much larger than what is required to heat the ICM (e.g., \citealt{Birzanetal2004}), we conclude that {{{{{ under these assumptions }}}}} turbulence cannot supply enough power to heat the ICM against radiative cooling. {{{{{ However, we note that the time scale of $t_{\rm AGN} \approx 10 \myr$ is highly uncertain, and that over a very long time that is much longer than $t_{\rm diss}$, turbulent dissipation and AGN heating might balance, such that turbulent dissipation can contribute to ICM heating. }}}}}   
 }}}} 
Overall the dissipation time is too long to account for pure turbulent heating. 

We conclude from this short discussion that the turbulence cannot be an important heating process in these clusters. The present conclusion is opposite to the conclusion of \cite{Lietal2020}. 
As we claimed in earlier papers (e.g., \citealt{HillelSoker2017b, HillelSoker2018}), the interaction of the jets and the bubbles they inflate with the ICM does drive turbulence, but it is a byproduct of many vortexes that this interaction forms, and not the major heating process.  To better illustrate this, we turn to analyse our earlier 3D hydrodynamical simulations.  

\section{THE NUMERICAL SCHEME}
\label{sec:numerics}

We present the flow structure of a 3D hydrodynamical simulation from   \cite{HillelSoker2016}, which we also analysed in \cite{HillelSoker2017b}. 
We describe here only the essential details of the numerical scheme (more information is in these two papers).  
   
We used the code {\sc pluto} \citep{Mignone2007} and simulated the octant, $x>0$, $y>0$ and $z>0$, where we take the $z$ axis along the jet's axis. The highest resolution of this adaptive mesh refinement grid is $\approx 0.1 \kpc$. We injected the jet from a circle $x^2 + y^2 \leq r_{\rm j}^2=(3 \kpc)^2$ in the $z = 0$ plane. The jet has a half-opening angle of $\theta_{\rm j} = 70^\circ$, and an initial velocity of $v_{\rm j} = 8200 \kms$ (for an observational support for wide and slow jets see, e.g., \citealt{Aravetal2013}). The power of the two jets together (as there is an opposite jet that we did not simulate) is
$\dot E_{2{\rm j}} = 2 \times 10^{45} \erg \s^{-1}$, and the mass loss rate in the two jets is $ \dot{M}_{2{\rm j}} = {2 \dot E_{2{\rm j}}}/{v_{\rm j}^2} = 94 M_{\odot}~\yr^{-1}$. The jet is intermittent, with an activity time period of $10 \myr$, namely, active phases at $t=0-10 \myr$, $20-30 \myr$ and so on, and an off time period of $10 \myr$, namely, the off periods are $t=10-20 \myr$, $30-40 \myr$, and so on.

The initial (at $t=0$) temperature of the ICM is $T_{\rm ICM} (0) = 3 \times 10^7 \K$, and the initial density profile is (e.g., \citealt{VernaleoReynolds2006})
\begin{equation}
\rho_{\rm ICM}(r) = \frac{10^{-25} \g \cm^{-3}}{\left[ 1 + \left( r / 100 \kpc \right)
^ 2 \right] ^ {3 / 4}}.
\end{equation}
The simulation includes a gravity field that maintains the gas at hydrostatic equilibrium before we inject the jets, and radiative cooling. 
     
\section{NUMERICAL FLOW STRUCTURE}
\label{sec:flow}

\subsection{Vortex scales}
\label{subsec:vortices}

\begin{figure}
\includegraphics[width=0.45\textwidth]{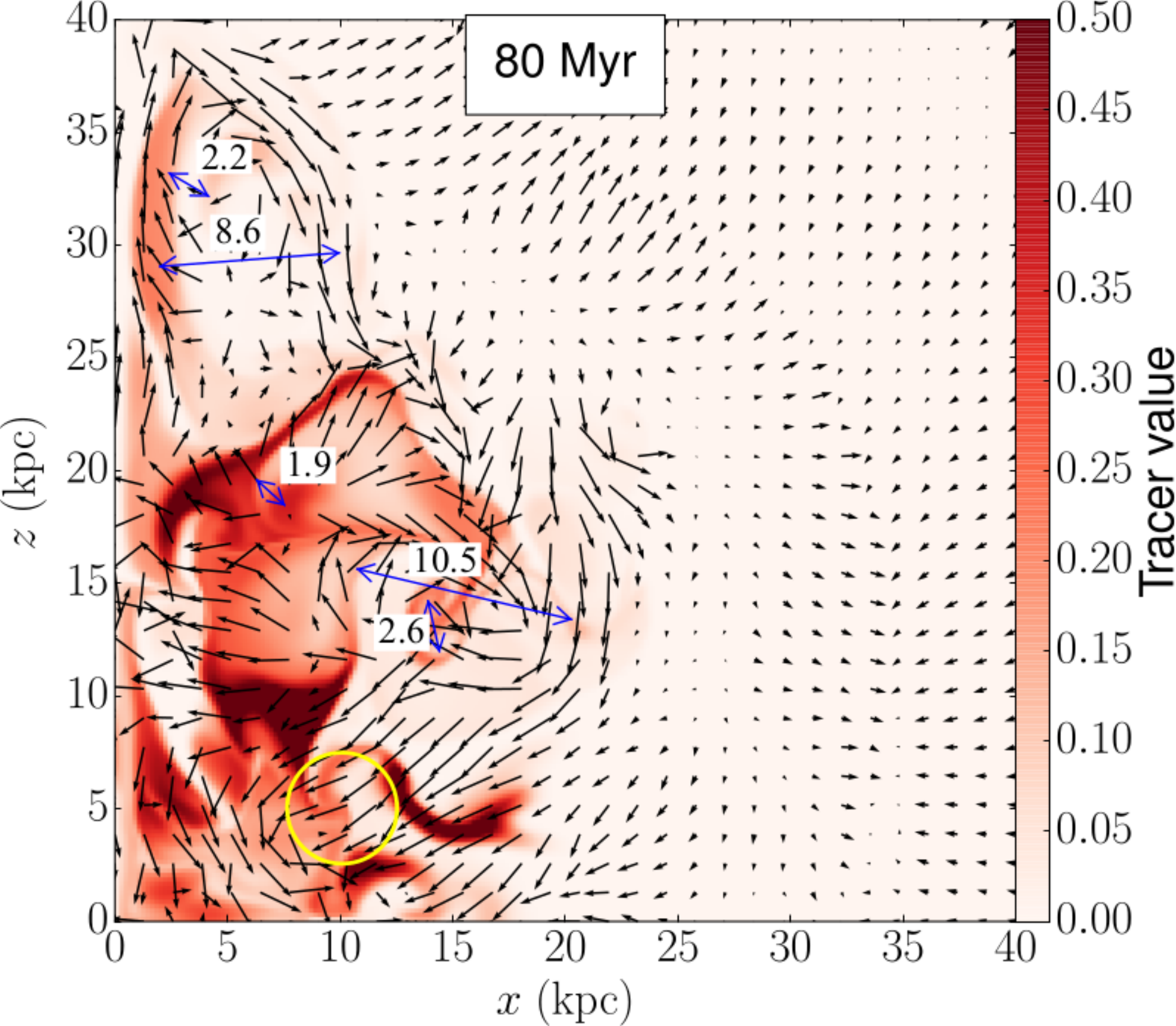}
\caption{The flow structure and the concentration of tracer A in the $y=0$ meridional plane at $t=80 \myr$.
At $t=0$ tracer A was inside a torus with a circular cross section having a radius of $r_{\rm tr}=2.5\kpc$ and centred at $(\varpi_c,z_c)_{\rm tr,A}=(10,5)\kpc$, where $\varpi=(x^2 + y^2)^{1/2}$. We mark this cross section by a yellow circle.
The longest velocity vectors correspond to velocities of $v \ge  400 \kms$.
Namely, we mark velocities of $>400 \km \s^{-1}$ with the same arrow length as for $400 \kms$. Blue double-headed arrows mark the distances, in kpc, between tracer segments with largely different velocity directions. }
 \label{fig:TracerA}
\end{figure}

The vortices that the jet-ICM interaction forms play a significant role  by mixing hot bubble content with the ICM (section \ref{sec:intro}). First we present the flow structure that reveals vortexes in one case from  \cite{HillelSoker2016}, that we also analysed in \cite{HillelSoker2017b} as tracer A. {{{{ Tracer A is frozen-in to the gas that at $t=0$ was inside a torus with a circular cross section having a radius of $r_{\rm tr}=2.5\kpc$ and centred at $(\varpi_c,z_c)_{\rm tr,A}=(10,5)\kpc$, where $\varpi=(x^2 + y^2)^{1/2}$ (a yellow circle in Fig. \ref{fig:TracerA}).  }}}}
Namely, the torus is parallel to the $x-y$ symmetry plane and its axis is the $z$ axis.
Fig. \ref{fig:TracerA} presents the flow structure in the $y=0$ meridional plane at $t= 80 \myr$. The colour coding depicts the concentration of a tracer A. A tracer is an artificial flow quantity that is frozen-in to the flow, and therefore it represents the spreading and mixing with time of the original parcel of gas. The initial value of the tracer inside the original volume is  $\xi (0) = 1$, and it is $\xi (0) = 0$ outside that volume. When the traced gas mixes with the ICM that started outside the original volume of the tracer or with the jets' material, its value drops to $0 < \xi(t) < 1$.

Figs.~\ref{fig:TracerA} demonstrates the following flow properties. (1) A complicated flow structure that the vortexes form. (2) The vortexes spread the tracer-gas over a large volume. (3) The vortexes span a large size range.

With the resolution we have it is impossible to resolve vortexes with diameters much less than about $1 \kpc$. The bubble size that the jet inflates (about the diameter of a sphere of the same volume as the bubble) is $D_{\rm bub} \simeq 20 \kpc$ \citep{HillelSoker2016}. We get here vortexes that are an order of magnitude smaller. One might imagine that the still narrow jet near the center might form small vortexes. Nonetheless, there are small vortex far from the center. 
  
\begin{figure}
\hskip -1.0cm 
\includegraphics[width=0.57\textwidth]{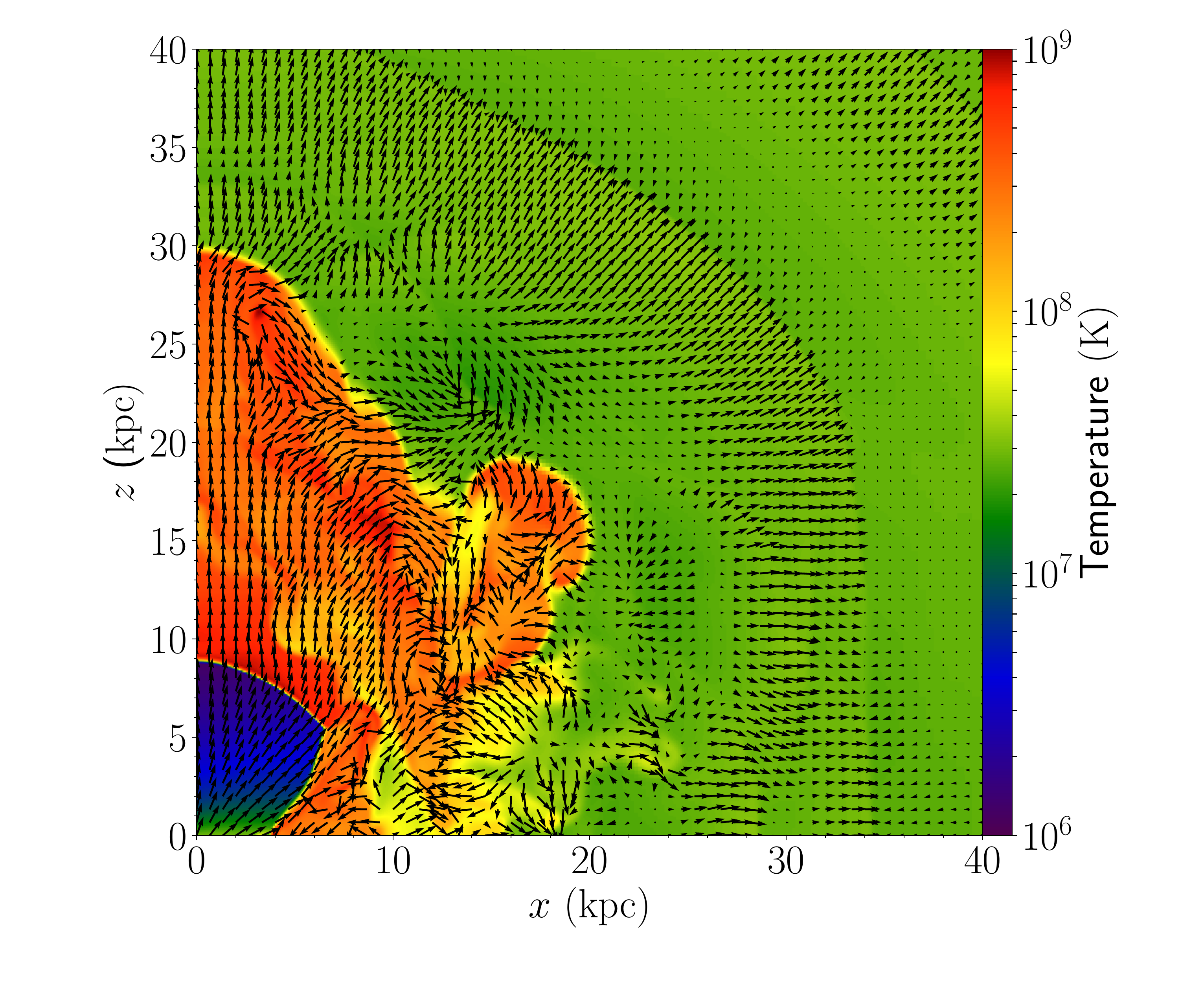}
\caption{Flow velocity and temperature maps at $t=44 \myr$ in the $y=0$ meridional plane. The temperature scale is according to the colour bar on the right. The length of each arrow is proportional to the velocity up to $v=150 \km \s^{-1}$. For any higher velocity the length of the arrow is as that of $v=150 \km \s^{-1}$. {{{{ The different zones in this meridional plane are as follows. The blue (very cool) region around the center is the freely expanding jet that suffers adiabatic cooling. It is surrounded by the hot (in red) post-shock jet's gas. The sharp boundary between the blue and red regions, which has a thin yellow line in this figure, is the shock wave of the jet. The large outer volume in green is the ICM before heating by mixing starts. The complicated regions between the red and green parts that are yellow-green, yellow, or yellow-red are regions where mixing of hot bubble gas (in red) with the ICM (in green) took place. }}}} }
 \label{fig:TempVelocity}
\end{figure}

To further analyse the flow structure we examine the temperature of the different flow zones in the ICM and in the bubble. In Fig. \ref{fig:TempVelocity} we present the temperature and the velocity at $t=44 \myr$. The length of each arrow is proportional to the velocity up to $v=150 \km \s^{-1}$. Any velocity of  $v > 150 \km \s^{-1}$ is represented by an arrow with the same length as for $v = 150 \km \s^{-1}$. {{{{ This way we emphasise the slow gas that is the focus of this study. Most of the gas that moves at higher velocity is in the pre-shocked jets (in \citealt{HillelSoker2016} we present more detailed velocity maps). }}}}
This figure shows that turbulence of different scales develops in the postshock region of the jet, in particular in the mixing zones with the ICM.

\begin{figure}
\hskip -1.2cm 
\includegraphics[width=0.57\textwidth]{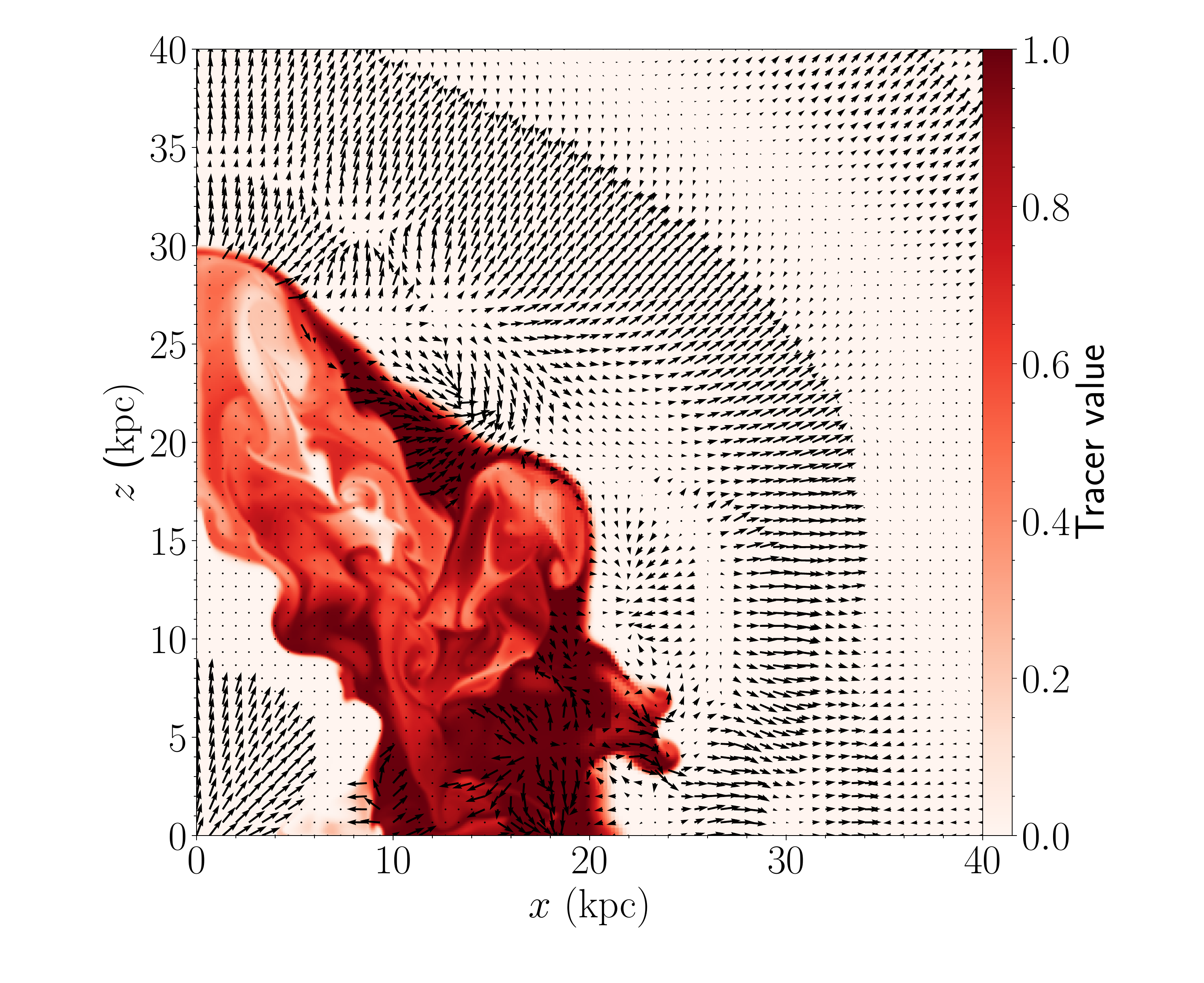}
\caption{Flow structure and the concentration of tracer C in the $y=0$ meridional plane at $t=44 \myr$. We present velocity arrows only for gas with a temperature of $T< 6 \times 10^7 \K$, to avoid hot bubble gas. At $t=0$ the tracer C was inside a sphere of radius $15 \kpc$ centred on the center of the grid (one octant). Velocity arrows as in Fig. \ref{fig:TempVelocity}. }
 \label{fig:Tracer15kpc}
\end{figure}

In Fig. \ref{fig:Tracer15kpc} we present the flow structure only of gas that has a temperature of $T<6 \times 10^7 \K=2 T_{\rm ICM}(0)$, so that we avoid hot bubble gas. Due to its adiabatic cooling, the velocity of the pre-shock jets is also in that map (near the center). In this figure we also present the distribution of tracer C, a tracer that is frozen to the gas that at $t=0$ was inside a sphere of radius $15 \kpc$ centred on the center of the grid (one octant).    
The tracer reveals a very complicated structure, with many small vortexes in the hot regions (where there are no arrows). The cooler regions also have a very complicated flow structure, with vortexes with sizes that span an order of magnitude. 
  
The conclusion from the results of this subsection is that the jet-ICM interaction can directly excite small-scale turbulence. Namely, the cascade from large scales to small scales accounts for only a fraction of the turbulent power at small scales in the ICM. We further show this in the next subsection. 

\subsection{No time to dissipate the large eddies}
\label{subsec:cascade}

\begin{figure}
\centering
\vskip -6.1 cm
\includegraphics[width=0.5\textwidth]{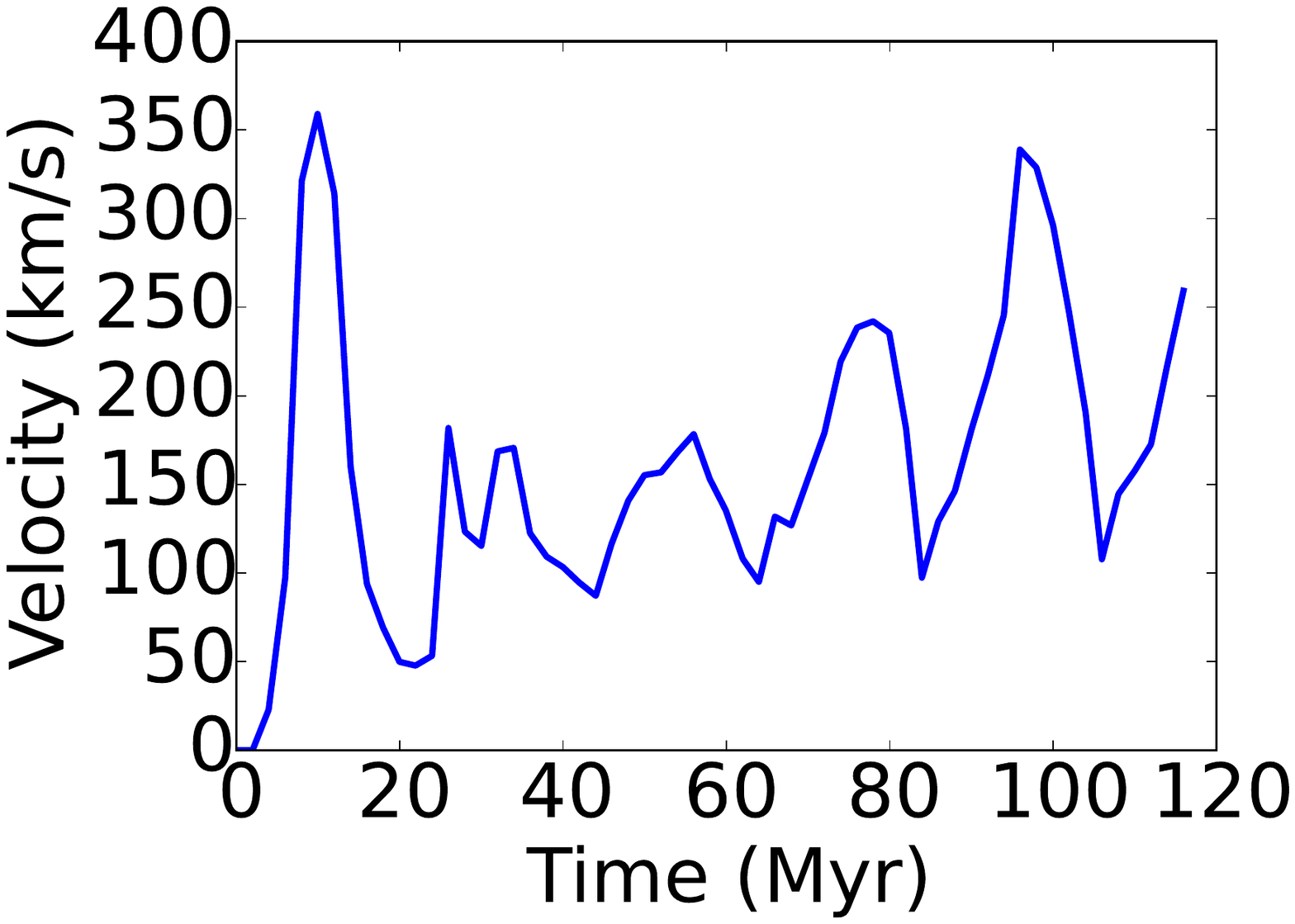}
\vskip -1.0 cm
\caption{The line-of-sight RMS velocity (numerical velocity dispersion; eq. \ref{eq:Vtracer}) of  tracer A, i.e., for numerical cells that include some tracer A (see Fig. \ref{fig:TracerA}), and only if the temperature in the cell is $T<4.5 \times 10^7 \K$ (so that it does not include hot shocked- jet's gas). }
\label{fig:Vtracer}
\end{figure}

The small vortexes (eddies) cannot come from the large vortexes by dissipation as there is no time for that. To show that, we use Fig. \ref{fig:Vtracer} that we taken from \cite{HillelSoker2017b}. 
In that earlier study we used this figure to show that the velocity dispersion of the ICM is similar in values to what observations with Hitomi show for the Perseus cluster \citep{Hitomi2016}. 

The velocity that Fig. \ref{fig:Vtracer} presents is the line of sight root mean square (RMS) velocity, which we termed numerical velocity dispersion, of all cells that contain even a little tracer A and also have a temperature of $T<4.5 \times 10^7 \K$. 
This velocity is 
\begin{equation}
\sigma_{\rm n} = \frac{\sqrt{<v^2>}}{\sqrt{3}} = \frac {1}{\sqrt{3}} \frac{2 E_{\rm k,tr}}{M_{\rm tr}}, \quad {\rm for} \quad T<4.5 \times 10^7 \K , 
\label{eq:Vtracer}
\end{equation}
where $E_{\rm k,tr}$ and $M_{\rm tr}$ are the kinetic energy and mass, respectively.

Fig. \ref{fig:Vtracer} shows two relevant properties to the present study. 
First it shows the dispersion velocity, that most of the time is $\sigma_n \la  250 \km \s^{-1}$. The turnover time of a vortex of size $L_{\rm max} = 10 \kpc$ is then $t_m \simeq L_{\rm max}/\sigma_n \ga 40 \myr$. The dissipation time is few times the turnover time, which is longer than the $t=80 \myr$ time of Fig. \ref{fig:TracerA} and the $t=44 \myr$ time of Figs. \ref{fig:TempVelocity} and \ref{fig:Tracer15kpc}. 

As well, Fig. \ref{fig:Vtracer} shows that the general dispersion velocity increases with each jet-launching episode. This shows that the energy has no time to dissipate, and that a different heating process is responsible for most of the ICM in cooling flows. 

From this numerical simulation and others, \cite{HillelSoker2016} found that only $\approx 20\%$ of the jet's kinetic energy is channelled to shock waves, sound waves, and a global flow. {{{{ This is compatible with the calculation of \cite{Formanetal2017} that in the Virgo cluster the shock carries $\approx 22 \%$ of the AGN energy. }}}}   Namely, heating by mixing, \cite{HillelSoker2016} concluded,  is the main heating process as about $80 \%$ of the jet's energy is channelled to heating the ICM by mixing. 
\cite{HillelSoker2017b} used this simulation to find that the numerical velocity dispersion is $\approx 100-250 \km \s^{-1}$, similar to the line-of-sight velocity dispersion of $164 \pm 10 \km \s^{-1}$ found by Hitomi in Perseus \citep{Hitomi2016}.

\section{The numerical velocity structure function}
\label{sec:VSF}

We now examine the numerical velocity structure function before the large vortexes have time to cascade down. We proceed as follows. 
\newline
(1) We take the flow at $t=44 \myr$, a time that ensures no significant cascade of the large turbulent eddies, since a typical cascade time is $t_{\rm diss} > t_m \simeq 10 \kpc / 100 \km \s^{-1}=100 \myr$. 
\newline 
(2) We interpolate the numerical adaptive mesh refinement (AMR) grid (where cells have different sizes) to a grid of $64 \times 64 \times 64$ cells, where all cells have the same size.  
\newline
(3) We mirror the octant grid about the planes $x = 0$, $y = 0$ and $z = 0$, so that we have a grid that covers all space around the center. 
\newline
(4) To avoid outer regions that the jets did not influence yet because of the short simulation time of $44 \myr$, we limit the volume we analyse to the ICM inside the ellipsoid 
$(x^2 + y^2)/(33\kpc)^2 + z^2/(39\kpc)^2 = 1$. 
{{{{ We term this the large-volume structure function. To examine the sensitivity to the volume we use,  we also calculate the numerical  velocity structure function for a smaller region that includes only regions close to the edge of the bubble. We take for the outer boundary of the regions of the small-volume structure function the surface $(x^2 + y^2)/(28\kpc)^2 + z^2/(35\kpc)^2 = 1$, which has its outer boundary at about half the distance from the bubble edge compared with that of the large volume above. }}}}
\newline
(5) To avoid the hot bubble gas we consider only gas with a temperature at or below the initial ICM temperature, i.e., we consider only ICM gas with $T< T_{\rm ICM}(0)=3 \times 10^7 \K$. We exclude the fast pre-shock jet gas (it is cold because of adiabatic cooling) by avoiding gas with velocities of $v> 10^3 \kms$. 
\newline
(6) For each pair of two cells that obey the above criteria, we record the distance $L_i$ and velocity difference $\vert \delta v_i \vert$ between the two cells. 
\newline
(7) We divide the pairs according to the distances $L_i$ in bins of $\Delta L = 0.1 \kpc$, and calculate the average velocity within each distance bin and obtain the velocity structure function $V_p(L) \equiv \langle \vert \delta v \vert\rangle$ as function of $L$. 
{{{{ For comparison, the largest cell size in the region we analyse is $0.2 \kpc$, which is twice as large as the smallest cell size in the entire numerical grid. }}}}

\begin{figure}
\centering
\includegraphics[width=0.5\textwidth]{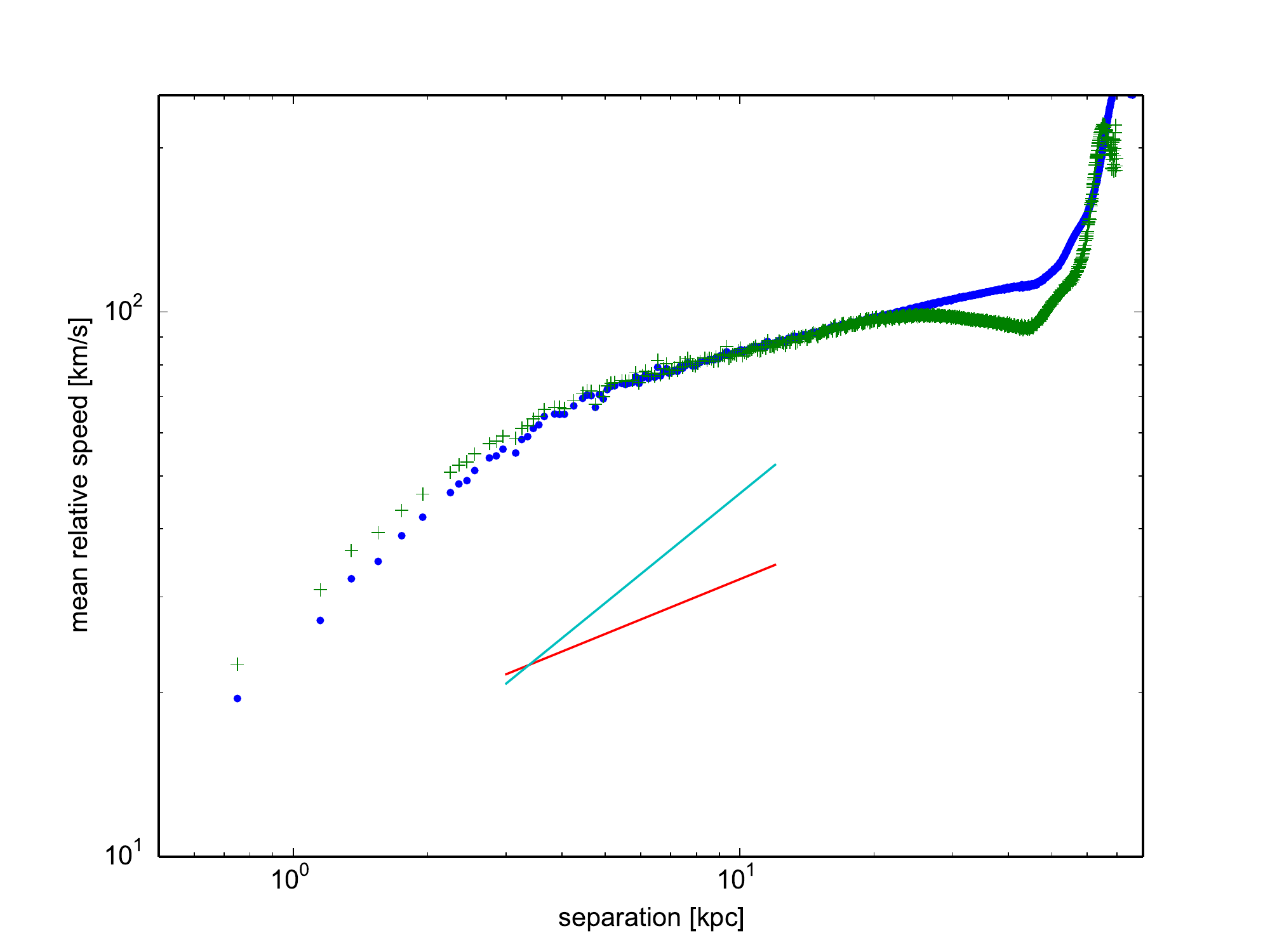}
\caption{The velocity structure function $V_p (L)$ of the ICM gas with a temperature of $T< T_{\rm ICM}(0)=3 \times 10^7 \K$. $V_p (L)$ is the average relative speed between two cells in a pairs, where the value at $L$ is the average over all pairs of cells with a distance between the two cells in the range of $L-0.05 \kpc$ to $L+0.05 \kpc$. 
{{{{ The blue-dots line is for the large-volume structure function calculated in a volume with an outer boundary at $(x^2 + y^2)/(33\kpc)^2 + z^2/(39\kpc)^2 = 1$, while the green-`$+$' line is for the small-volume structure function calculated inside the surface $(x^2 + y^2)/(28\kpc)^2 + z^2/(35\kpc)^2 = 1$. These volumes do not include the hot bubble itself. }}}} The two straight lines have a slope of $1/3$ (red) and $2/3$ (pale blue). }
\label{fig:VSF}
\end{figure}

We present the numerical velocity structure function in Fig. \ref{fig:VSF}. {{{{ The differences between the large-volume structure function (blue dots) and the small-volume structure function (green-`$+$' symbols) are very small, in particular in the relevant range. We are not sensitive to the choice of the region when calculating the numerical velocity structure function. }}}} From this figure we learn that the process of bubble inflation excites turbulence over more than an order of magnitude in scale, much before the large eddies, $L \simeq 10-20 \kpc$, have time to cascade and form small eddies, $L \la {\rm few} \times \kpc$. This strengthens the results of section \ref{sec:flow}. We see that some parts are steeper, $45 \la L $ and $ L \la 5 \kpc$, and some are shallower, $5 \la L \la 45 \kpc$, than the classical Kolmogorov expectation (a slope of 1/3).
 
We do not take the velocity structure function that we obtain here to be universal. We only claim that as jets inflate bubbles they excite turbulence with eddies with sizes over more than an order of magnitude. The velocity structure function depends on the properties of the jets and the preexisting weak turbulence in the ICM. It might well be that jets can induce a turbulence where at all scales the velocity structure function is steeper than $1/3$, as \cite{Lietal2020} infer for three clusters.  
   
\section{SUMMARY}
\label{sec:summary}

The conclusion of this short study is that the jet-ICM interaction drives vortexes (turbulent eddies) in a large range of scales (Figs. \ref{fig:TracerA}-\ref{fig:Tracer15kpc}), that in turn drive the turbulence in the ICM with eddies of over more than an order of magnitude in size (Fig. \ref{fig:VSF}). We argue, therefore, that the dissipation of the large turbulent eddies is not the main process that determine the velocity structure function of the optical filaments that \cite{Lietal2020} find, but rather the excitation of the turbulence by the jet-ICM interaction. Indeed, the turbulent properties do not allow for an efficient heating of the ICM in these three cooling flow clusters (section \ref{sec:Turbulence}). 
 
We did not build our earlier 3D hydrodynamical simulations \citep{HillelSoker2016} to study the velocity structure function of cold filaments. We encourage the study of velocity structure functions in 3D hydrodynamical simulations of jets that inflate bubbles in the ICM. For that, the simulations should replace the simple tracer by a volume that has a gas with a somewhat lower temperature than that of the ICM. After a long time the gas will cool and form filaments. The velocity structure function of these numerical filaments can be compared with the velocity structure functions that \cite{Lietal2020} deduce from observations, in cases with strong viscosity, where cascade down is rapid, and in cases with very small viscosity where cascade down is negligible. Our predction is that the jets-ICM interaction by itself can explain most (but not all) of the properties of the velocity structure functions. 

{{{{ We thank an anonymous referee for detailed comments that improved the manuscript. }}}} This research was supported by the Prof. Amnon Pazy Research Foundation. For computational resources we acknowledge the LinkSCEEM/Cy-Tera project, which is co-funded by the European Regional Development Fund and the Republic of Cyprus through the Research Promotion Foundation.

\label{lastpage}
\end{document}